\documentclass{PoS}
\usepackage{xcolor}
\usepackage[normalem]{ulem}
\title{Searching for High-Energy Neutrino Emission from TeV Pulsar Wind Nebulae}

\ShortTitle{Searching for High-Energy Neutrino Emission from TeV Pulsar Wind Nebulae}

\author{
The IceCube Collaboration\footnote{For collaboration list, see PoS(ICRC2019) 1177.}\\
{\itshape \href{http://icecube.wisc.edu/collaboration/authors/icrc19_icecube}{http://icecube.wisc.edu/collaboration/authors/icrc19\_icecube}}\\
E-mail: \email{qliu@icecube.wisc.edu}, \email{akheirandish@icecube.wisc.edu}
}

\abstract{
Pulsar wind nebulae (PWNe) are main gamma-ray emitters in the Galactic plane. Although the leptonic scenario is able to explain most PWNe emission well, a hadronic contribution cannot be excluded. High-energy emission raises the possibility that gamma-rays are hadronically produced which inevitably leads to the production of neutrinos. We report a stacking analysis to search for neutrino emission from 35 PWNe that are very-high-energy gamma-ray emitters and the results using 9.5 years of all-sky IceCube data. In the absence of any significant correlation, we set upper limits on the total neutrino emission from those PWNe and constraints on the hadronic component.\\

\vspace{4mm}
{\bfseries Corresponding authors:}
 \speaker{Qinrui Liu}$^{1}$, {Ali Kheirandish}$^{1}$\\
{$^{1}$ \itshape Wisconsin IceCube Particle Astrophysics Center (WIPAC) and Department of Physics, University of Wisconsin-Madison, Madison, WI 53706, USA}\\
}

\FullConference{36th International Cosmic Ray Conference -ICRC2019-\\
		July 24th - August 1st, 2019\\
		Madison, WI, U.S.A.}

\begin{document}
\section{Introduction}\label{sec:info}
Galactic cosmic rays (CRs) are believed to be dominant at least up to the knee of the CR spectrum. The interactions of those particles are expected to generate photons and neutrinos from decays of secondary pions which can reach energies of hundreds of TeV \cite{Ahlers:2015moa}. As very high energy gamma-rays could also originate in leptonic scenarios, the smoking gun for identification of a Galactic cosmic accelerator relies on identifying a high-energy neutrino source. 

Since the discovery of high-energy neutrinos of astrophysical origin \cite{Aartsen2013e,Aartsen:2014gkd}, IceCube has conducted analyses searching for the sources of cosmic neutrinos and Galactic objects have been considered as potential sources. If the production of high-energy gamma-rays in the Galaxy is hadronic, they would be accompanied by high-energy neutrinos. Potential sources of the high-energy neutrinos in the Galaxy should in principle be predefined from the observation of high-energy gamma-ray emission. Imaging Air Cherenkov Telescopes (IACTs) and water Cherenkov telescopes provided a good and comprehensive view of the Galactic plane at high energies, for instance, surveys by H.E.S.S. and HAWC \cite{H.E.S.S.:2018zkf, Abeysekara:2017hyn}. Interestingly, the majority of these objects were found to be pulsar wind nebulae.

Pulsar wind nebulae (PWNe) are diffuse nebulae confined inside supernova remnants (SNR) that are powered by pulsar winds generated by the highly spinning and magnetized pulsars in the center. High-energy photons are believed to be primarily emitted by electron-positron pairs accelerated in pulsar winds. Leptonic scenarios are able to explain photons from radio wavelengths to TeV energies, in which low-energy emission is dominated by the synchrotron process and the inverse Compton scattering of synchrotron photons becomes dominant at high energies \cite{DiPalma:2016yfy}. However, the presence of hadrons coexisting with leptons is still uncertain and to date cannot be excluded by theory or observations. Hadronic mechanisms are discussed for example in \cite{DiPalma:2016yfy,Cheng:1990au, Bednarek:1997cn,Bednarek:2003cv, Amato:2003kw}.

\section{Analysis}\label{sec:ana}
We consider thirty-five PWNe which have been observed with > 1 TeV gamma-ray emission by high-energy gamma-ray telescopes such as HAWC, H.E.S.S., MAGIC and VERITAS. We use an unbinned maximum likelihood method \cite{Braun:2008bg} to perform a stacking search for neutrino emission from these sources. Stacking potential sources together, which means combining signals from different sources, can improve the sensitivity of the analysis \cite{Achterberg:2006ik}. After modifying the likelihood function in \cite{Braun:2008bg} for a stacking search, it can be written as

\begin{equation}
\label{eqn:llhfunc}
\mathcal{L}(n_{s}, \gamma_{s}) = \prod^{N}_{i}(\sum^{M}_{j}\omega_{j}\frac{n_{s}}{N}S_{i}^{j}+(1-\frac{n_{s}}{N})B_{i}),
\end{equation}
where $n_{s}$ is the number of signal events, $N$ is the total number of neutrino events and $M$ is the number of sources. $S_{i}^{j}$ is the signal probability density function (PDF) which corresponds to the $i$th event with respect to the $j$th source. The normalized weight, $\omega_{j}$, determines the relative normalization of the signal PDF from source $j$. $B_{i}$ is the background PDF. The PDFs are composed of a spatial term and an energy term. For the signal PDF, we have $S_{i}^{j}=S^{S}(x_{j}, x_{i}, \sigma_{ij})\times S^{E}(E_{i}, \gamma_s)$. The spatial clustering of signal events is modeled as a two-dimensional Gaussian distribution. The width of the spatial PDF, $\sigma_{ij}$, which represents the effective angular uncertainty $\sigma_i$ of event $i$ and the angular extension $\sigma_j$ of source $j$, is defined as $\sigma_{ij}=(\sigma_{i}^2+\sigma_{j}^2)^{1/2}$. An unbroken power-law spectrum is assumed for calculating the energy term of the signal PDF. In order to avoid bias, we set the spectral index $\gamma_s$ as a generic parameter for all sources instead of using the measured index for each source from gamma-ray observations. The signal we are interested in is an excess of neutrinos in the directions of the sources where an isotropic background is expected. The background is constructed from the data itself by randomizing the right ascension of events. The null hypothesis presumes no signal-like event, i.e., $n_{s}=0$. The test statistic (TS) is defined by a log-likelihood ratio $\mathrm{TS}=2\log(\mathcal{L}(\hat{n}_{s},\hat{\gamma}_{s})/\mathcal{L}(n_{s}=0))$ in which $(\hat{n}_{s},\hat{\gamma}_{s})$ are the best-fit values. 

\begin{figure}[!htb]
\vspace*{-5pt}
\centering
\includegraphics[width=0.6\linewidth]{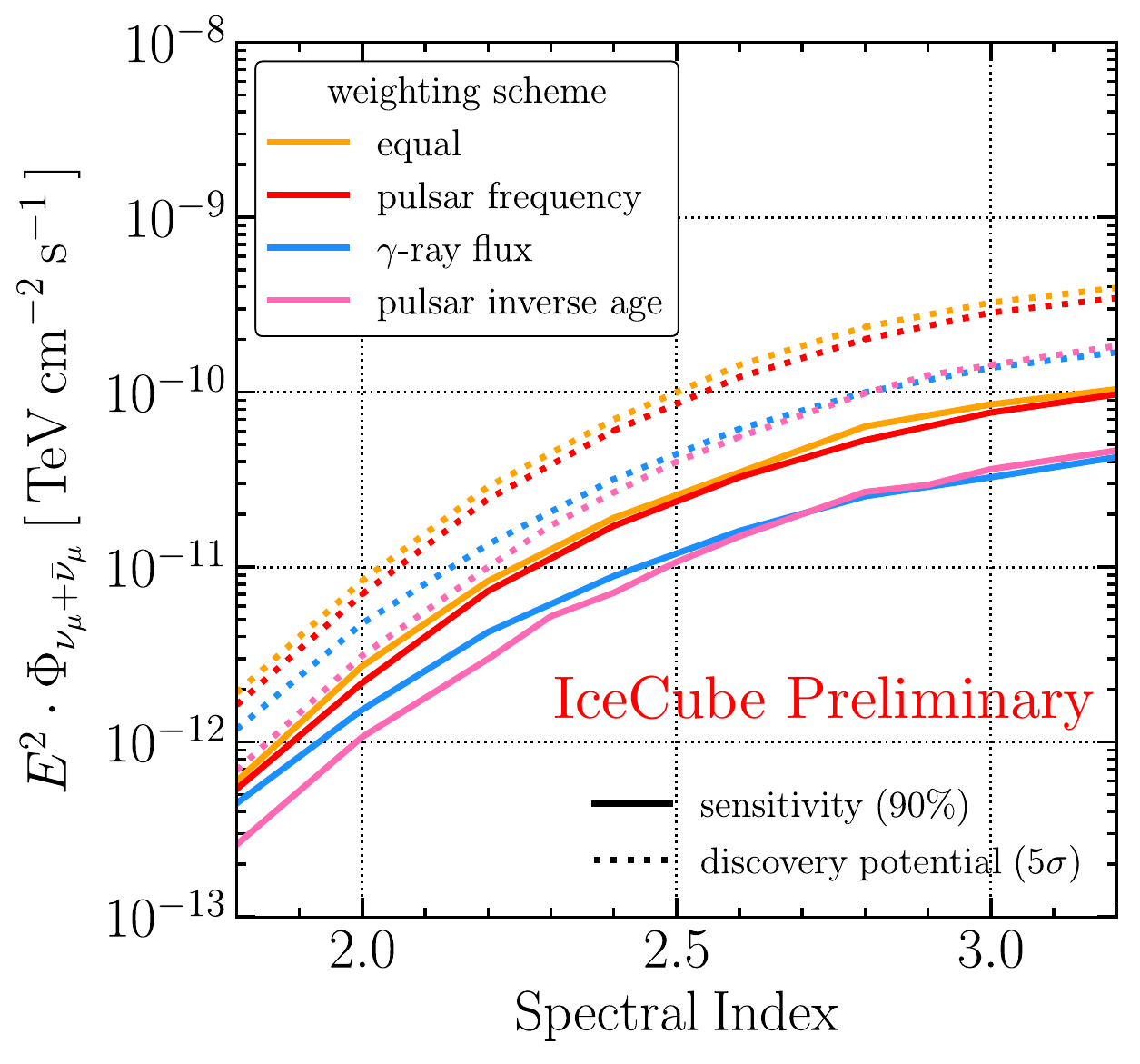}
\caption{Sensitivities (90$\%$ CL) and 5$\sigma$ discovery potentials of different weighting schemes as a function of the spectral index for an unbroken power-law spectrum injected from sources at 1 TeV.}
\label{fig:sens_disc}
\end{figure}

Weights applied to each source can be from theory or observational measurements in order to test a specific hypothesis. Here, four different hypotheses are tested by incorporating different weighting schemes:

{\bf{Equal weighting --}} No preference of neutrino emission probability is given to any source and all sources are treated equally with $\omega_j=1/M$.

{\bf{Flux weighting --}} The assumption is that a plausible high-energy neutrino emission is proportional to the high-energy gamma-ray emission from each source. If true, this implies that the high-energy gamma rays are either partially or completely of hadronic origin. For sources in the Northern Sky, spectral measurements of VERITAS and MAGIC are used; while for sources in the Southern Sky, H.E.S.S. is used as it is more sensitive in this region and HAWC observations are used for Geminga and 2HWC J0700+143. In this analysis, the gamma-ray flux $\phi$ at 1 TeV is incorporated, i.e. for source $j$, $\omega_j=\phi_j/\sum_{k}^{M}{\phi_k}$.

{\bf{Frequency weighting --}} The energy carried by the pulsar wind for the acceleration is taken from the rotational energy of the pulsar as it emits radiation, which results in the spin-down of the pulsar \cite{Gaensler:2006ua}. Thus, the period of the pulsar is an important measure of how energetic the pulsar is. Using frequencies $f$ as weights, $\omega_j=f_j/\sum_{k}^{M}{f_k}$, faster spinning sources are preferred in this scenario.

{\bf{Age weighting --}} This parameter is used to estimate the true age of a pulsar under assumptions that the initial spin was much faster than it is today and the energy loss is from magnetic dipole radiation \cite{Gaensler:2006ua}. Here, the inverse age $\tau^{-1}$ is used as the weight for each source, $\omega_j=\tau_j^{-1}/\sum_{k}^{M}{\tau_k^{-1}}$. This assumption prefers younger PWNe to be more energetic emitters.

In this analysis, we use the 9.5-year all-sky data collected by IceCube between April 2008 and November 2017. This includes seven years of data, already studied for neutrino point sources \cite{Aartsen:2016oji}, along with additional data for the period of May 2015 and 2017 \cite{Aartsen:2016lmt}. These periods differ in detector configuration, data-taking conditions, and event selections. To estimate the performance of the analysis, source emission is simulated to observe the detector response. Sensitivities (90$\%$ confidence level (CL)) and discovery potentials (5$\sigma$) for different weighting schemes discussed above are shown in Figure \ref{fig:sens_disc}. To simulate the neutrino emission, an unbroken power-law spectral shape is assumed. The projected sensitivity shows, as expected, that IceCube is more sensitive to sources with a harder spectrum. Also, if the neutrino emission is correlated with the observed high-energy gamma ray emission, IceCube would be more likely to see an excess of neutrinos.
%{\color{blue} talk about result, no evidence, %discuss each scenario, and how upper limit is %obtained}
\section{Results}\label{sec:results}
\begin{table}[!b]
\centering
\begin{tabular}{l|c|c|c|c|c|c|c}
weighting&TS&$\hat{n}_{s}$&$\hat{\gamma}_{s}$&p-value&$\Phi_{\nu_{\mu}+\bar{\nu}_{\mu}}^{90\%,\;E^{-2.0}}$&$\Phi_{\nu_{\mu}+\bar{\nu}_{\mu}}^{90\%,\;E^{-2.19}}$&$\Phi_{\nu_{\mu}+\bar{\nu}_{\mu}}^{90\%,\;E^{-2.5}}$ \\
\hline
Equal&0.81&40.43&3.84&22.58\%&3.91&11.6&44.5\\
Frequency&0.26&18.0&3.81&37.85\%&2.64&7.79&28.2\\
Flux&0.21&8.73&4.00&36.17\%&1.74&4.57&14.9\\
Age&0&0&-&-&1.07&2.82&10.7\\
\end{tabular}
\caption{Best-fits for TS, $n_{s}$ and $\gamma_s$ of the four different weighting schemes discussed in section \ref{sec:ana}. TS=0 is obtained under age weighting so that no other best-fit is given. The last three columns are upper limit constraints on the stacking flux with a 90\% CL. The first one has a power-law spectrum $E^{-2.0}$, the second has $E^{-2.19}$, and the last column follows $E^{-2.5}$. They are all normalized at 1$\;\rm{TeV}$ with units $10^{-12}\;\mathrm{TeV^{-1}cm^{-2}s^{-1}}$.}
\label{tab:Results}
\end{table}
%\vspace{-1pt}
The unbinned maximum likelihood analysis discussed in Sec \ref{sec:ana} is performed considering equal, gamma-ray flux, frequency and inverse age weightings. The results for each test are presented in Table \ref{tab:Results}. No significant excess of signal events was found. The largest excess was found in the equal weighting scheme, which yields a best-fit signal of 40.4 events with a pre-trial p-value of 0.22. 90\% CL upper limits on the total flux from TeV PWNe under each hypothesis can be obtained. For this purpose, we have assumed an unbroken power-law flux with three different spectral indices, which are $E^{-2}$, $E^{-2.19}$ and $E^{-2.5}$. The latter two correspond to the fitted spectrum of astrophysical muon neutrinos \cite{Haack:2017dxi} and the global fit analysis of IceCube data between 25 TeV and 2.8 PeV \cite{Aartsen:2015knd}, respectively. 

The Galactic component of the high-energy neutrino flux is constrained to $\sim14\%$ of the total flux at 1 TeV \cite{Aartsen:2017ujz}. Considering the high-energy neutrino flux reported in \cite{Haack:2017dxi}, the contribution of neutrino emission from TeV PWNe studied here to the total muon neutrino flux is less than $\sim4\%$, and less than $\sim2\%$ if considering the global fit flux reported in \cite{Aartsen:2015knd}. One should note that this limit is valid within the specific assumptions of this analysis regarding the weighting and selection of the sources and should not be applied or extended to other hypotheses.

\begin{figure}[!h]
\centering
\includegraphics[width=0.6\linewidth]{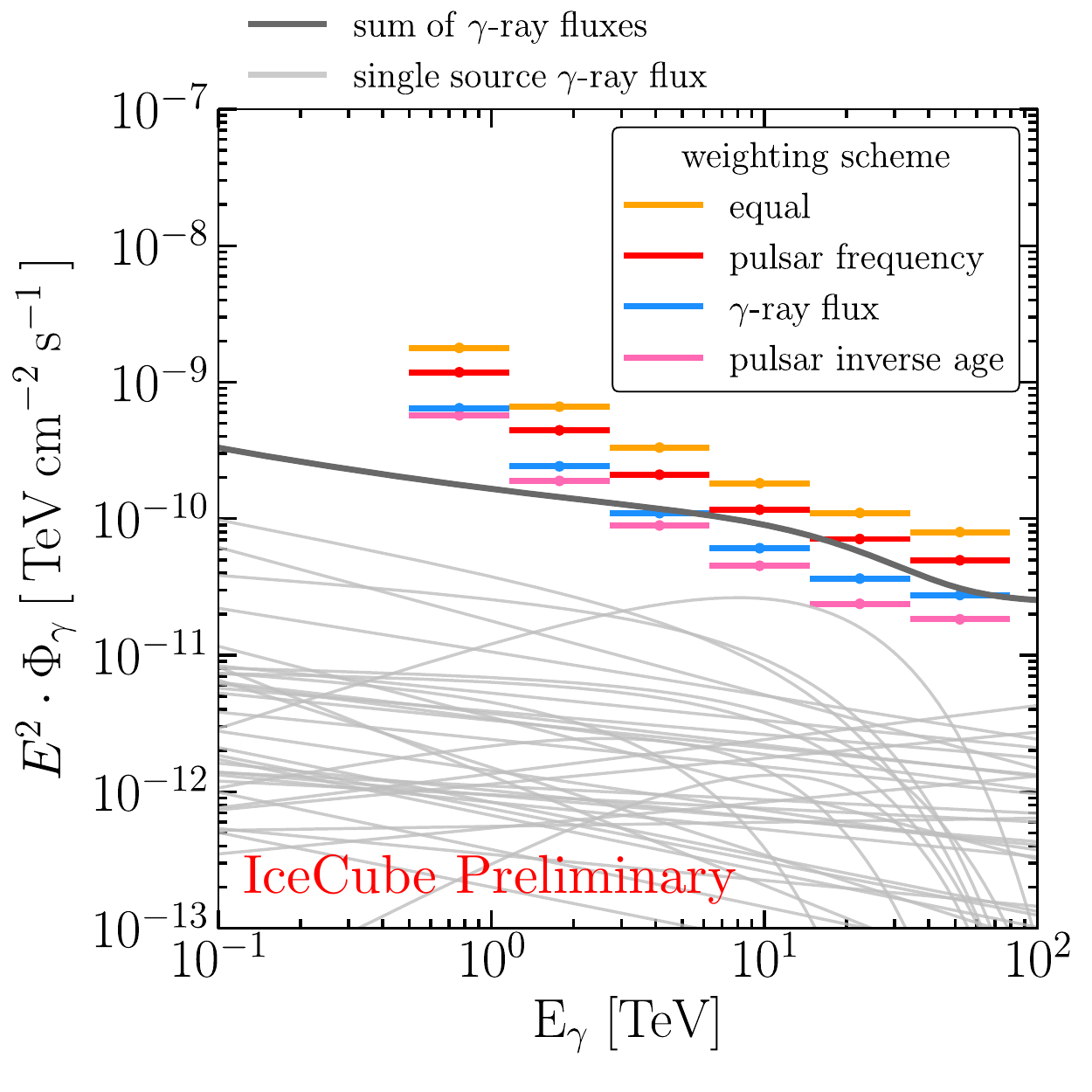}
\caption{Light gray lines are observed gamma-ray spectra from sources, and the dark gray line is the sum of those fluxes. Orange, red, blue, and pink steps are hadronic gamma-ray upper limits converted from 90$\%$ CL neutrino upper limits, and each color corresponds to a given weighting method. The spectrum of the upper limits shown here is $E^{-2}$. To avoid uncertainties from extrapolation, the energy only goes to 100 TeV in the plot.}
\label{fig:UL}
\end{figure}

One can use the upper limit on the neutrino flux to constrain the hadronic component of the observed high-energy gamma-ray flux with the relation described in \cite{Ahlers:2013xia}. We assume proton-proton interactions at the sources to convert neutrino fluxes to their gamma-ray counterparts. High-energy gamma ray flux measurements extend to tens of TeV while IceCube neutrinos reach energies of a few PeV. To avoid large uncertainties due to the extrapolation of the high gamma-ray flux, we calculate differential upper limits assuming an unbroken power-law spectrum and convert them to the gamma-ray flux upper limits and restrict them to 100 TeV. Figure \ref{fig:UL} shows differential upper limits for $E^{-2}$ spectrum for different hypotheses tests of this study compared to the cumulative flux of high-energy gamma-rays.

\section{Summary}\label{sec:discussion}
The main contributors to IceCube cosmic neutrinos are extragalactic, however, Galactic CR accelerators are guaranteed, although limited, to contribute to the total observed high-energy cosmic neutrino flux. In this study, we examined the possible neutrino emission from PWNe with TeV gamma-ray emission. Thirty-five sources were tested, and the stacking search for the high-energy neutrino emission preferred the isotropic arrival direction hypothesis. In the absence of a significant excess of neutrino events in the direction of these sources, we have set upper limits on the neutrino emission from these sources and potential hadronic component of the high-energy gamma-ray flux.

The stacking analysis presented here yields upper limits at the level of the total observed high-energy gamma ray emission. While the addition of more years of data with continuous operation of IceCube will improve the sensitivity of the search for Galactic sources of cosmic neutrinos, more accurate measurements of the very high energy gamma-ray flux by HAWC and future gamma-ray observatories such as CTA will shed more light on the nature of the emission and the shape of the spectrum.

% Set up the bibliography using BibTeX.
% Get references from inspirehep.net or NASA/ADS and put them in references.bib.
\bibliographystyle{ICRC}
\bibliography{references}

% Or, set up the bibliography manually, if you prefer to do things this way.
%
% \begin{thebibliography}{99}
%   \bibitem{Zoll:2015wcu}{{\bf IceCube} Collaboration, \pos{PoS(ICRC2015)1099} (2016).}
%   \bibitem{Peiffer:2017vsm}{{\bf IceCube-Gen2} Collaboration, \pos{PoS(ICRC2017)1052} (2018).}
%   \bibitem{Hussain:2019icrc_gw}{{\bf IceCube} Collaboration, \pos{PoS(ICRC2019)xyz} (these proceedings).}
%   \bibitem{Aartsen:2016nxy}{{\bf IceCube} Collaboration, M.~G.~Aartsen {et al.}, \emph{JINST} {\bf 12} (2017) P03012%
%   % optionally add arXiv ID here [{\tt astro-ph/1612.05093}]
%   .}
%   \bibitem{Waxman:1998yy}{E. Waxman and J. N. Bahcall, \emph{Phys. Rev.} {\bf D59} (1999) 023002.}
% \end{thebibliography}

\end{document}